# Quantum Number Density Asymmetries Within QCD Jets Correlated With $\Lambda_0$ Polarization

**Dennis Sivers**

Portland Physics Institute
4730 SW Macadam #101
Portland, OR 97239

Spin Physics Center
University of Michigan
Ann Arbor, MI 48109

**Abstract**

The observation of jets in a variety of hard-scattering processes has allowed the quantitative study of perturbative quantum chromodynamics (PQCD) by comparing detailed theoretical predictions with a wide range of experimental data. This paper examines how some important, <u>nonperturbative</u>, facets of QCD involving the internal dynamical structure of jets can be studied by measuring the spin orientation of $\Lambda_0$ particles produced in these jets. The measurement of the transverse polarization for an individual $\Lambda_0$ within a QCD jet permits the definition of spin-directed asymmetries for quantum number densities in rapidity space (such as charge, strangeness and baryon number densities) involving neighboring hadrons in the jet. These asymmetries can only be generated by soft, nonperturbative dynamical mechanisms and such measurements can provide insight not otherwise accessible into the color rearrangement that occurs during the hadronization stage of fragmentation process.

PACS: 12.83.-t, 13.87.-a, 13.80.+e, 14.20.Jn

**Introduction**

Powerful factorization theorems [1] allow the study of perturbative quantum chromodynamics, PQCD, in hard-scattering processes involving hadrons. A quantitative understanding of fundamental multi-jet processes representing the scattering of quarks and gluons currently provides the foundation for theoretical and experimental efforts to understand electroweak symmetry breaking and to search for possible physics mechanisms beyond the Standard Model. [2,3] The discussion presented here, however, concerns a quite different type of factorization also found in QCD. In contrast to the elaborate genesis of hard-scattering factorization that is now contained in multiple textbooks and summary reviews [4], this additional factorization property can be directly traced to a single influential paper by Kane, Pumplin and Repko [5], and so can aptly be named KPR factorization [6].

The property of KPR factorization recognizes that the basic result of Ref. [5],

$$A_N d\sigma(qq\uparrow \Rightarrow qq)/d\sigma(qq \Rightarrow qq) = \alpha_s(Q^2)\left(\frac{m_q}{Q}\right)f(\Theta_{CM}), \qquad (1)$$

where $A_N d\sigma(qq\uparrow \Rightarrow qq) = \frac{1}{2}[d\sigma(qq\uparrow \Rightarrow qq) - d\sigma(qq\downarrow \Rightarrow qq)]$ is the transverse spin asymmetry for quark-quark scattering. The existence of similar expressions quark-gluon and quark antiquark scattering requires that significant parity-conserving transverse single-spin asymmetries cannot be generated in PQCD processes involving light quarks because gauge interactions preserve quark helicities for light quarks. This does not mean, however, that such transverse spin observables are absent in QCD [7]. In fact, transverse spin asymmetries must exist in the full quantum field theory because of the spin-orbit dynamics required by the interplay of confinement and dynamic chiral symmetry breaking. Within the phenomenological study of hard-scattering processes, the asymmetries generated by such spin-orbit dynamics can be absorbed into $p_T$-dependent effective distribution functions (orbital distributions [8] or Boer-Mulders functions [9]) or into $p_T$-dependent fragmentation functions (Collins functions [10] or polarizing fragmentation functions [11]). Alternately, these nonperturbative dynamical mechanisms can be parameterized, within the overall framework of collinear factorization, by several specific, twist-3, local operators [12]. Either approach factorizes the spin-directed dynamics into a system that can be probed by hard scattering and allows perturbative QCD to be used in the study of significant aspects of nonperturbative dynamics.

A growing subfield of particle physics that can be designated transverse-spin physics depends directly on the existence of KPR factorization. The boundaries of this subfield can be identified by a set of conventions, the Trento Conventions [13], which allow unambiguous comparison of experiments and theoretical predictions involving transverse spin. This subject received a boost when it was pointed out by Heppelman, Collins and Ladinsky [14] that the quark transversity distributions, $\delta^T q(x)$, giving the momentum distributions of transversely polarized quarks in a transversely polarized proton as defined by Ralston and Soper [15] and renamed by Jaffe and Ji [16] could be measured in semi-inclusive deep inelastic scattering, SIDIS,

$$A_N d\sigma(lp\uparrow \Rightarrow l\pi X) \propto \delta^T q(x) \otimes \Delta^N D_{\pi/q\uparrow}(z), \qquad (2)$$

where $\Delta^N D_{\pi/q\uparrow}(z)$ is the Collins function that defines an asymmetry in the fragmentation of a transversely polarized quark. The level of both experimental and phenomenological progress in transverse spin physics has recently been quite high. For example, various sets of comprehensive phenomenological fits to experimental asymmetries in $e^+e^- \Rightarrow hadrons$, SIDIS and inclusive production in polarized hadron-hadron scattering have been published [17]. These fits have provided parameterizations for the transversity distributions of u and d quarks, for favored and disfavored Collins functions, and for orbital distributions of u and d quarks. The phenomenological fits have already provided considerable insight into significant nonperturbative mechanisms in QCD and into the internal dynamical structure of the proton. The initial studies have thus created new expectations and further experimental programs involving transverse spin physics have been approved with the hope for additional progress.

The subject of this paper provides evidence that the applications of KPR factorization in QCD dynamics to the study of hadronic processes can, in fact, be much broader than is indicated by existing phenomenological treatments described above. Because the mechanisms that can produce transverse spin asymmetries necessarily involve soft, nonperturbative dynamics, such observables provide a versatile tool for the study of the internal structure for a large variety of different subsystems of a scattering event. For example, consider the production of a $\Lambda_0$ particle within a QCD jet generated by a hard collision as illustrated in Fig. 1. For the purpose of this discussion, it does not matter whether the jet is found in $e^+e^- \Rightarrow hadrons$, lepton-nucleon, hadron-hadron or hadron-nucleus collisions but we will assume that the $\Lambda_0$ under consideration is detected in the "mid-rapidity" range of the jet and that its momentum transverse to the jet axis can be measured and is nonzero so that a plane can be defined containing the jet axis and the $\Lambda_0$ momentum. It is also assumed that the momenta of other hadrons within the jet can be measured. The point is that each individual $\Lambda_0$ particle has a spin direction and this spin direction can be measured by the angular dependence of its parity-violating weak decay $\Lambda_0 \Rightarrow p\pi^-$. The transverse component of the spin-direction for each $\Lambda_0$ particle then defines an orientation to the plane containing jet axis and $\Lambda_0$ momentum. We can use that orientation to specify parity-conserving spin-directed asymmetries in momentum correlations connecting the $\Lambda_0$ hyperon with other neighboring hadrons in the jet. It is convenient to define the resulting asymmetries in the $\Lambda_0$-hadron correlation functions in terms of rapidity-space quantum number densities.   Therefore, in this paper, we will introduce the basic application of KPR factorization to hadronization stage of jet fragmentation and briefly consider possible asymmetries in charge, strangeness and baryon number densities within a QCD jet and their correlation with the orientation of $\Lambda_0$ spin.  Obviously, the techniques discussed in this paper can be applied to other hyperons or anti-hyperons with detectable weak decays.  Taken together, a set of such measurements can be used to describe an ensemble of complex spin-directed dynamical mechanisms occurring in the jet fragmentation process within a finite rapidity range of the measured hyperon.

**Fragmentation Dynamics**

The application of the multiple constraints implied by the confinement of color charge and by the conservation of fundamental quantum numbers in the "fragmentation" of a fundamental QCD constituent (quark or gluon) produced in a hard-scattering process into a final state consisting of an ensemble of color-singlet hadrons provides one of the most interesting set of challenges in quantum field theory.  In the quantitative treatment of jet production in hard-scattering processes, the analysis of an event starts from the specific definition of a jet in terms of the resolution parameter appearing in a jet-finding algorithm [18].  In this way the multiparticle final state in transformed into a final state with only a few jets that can be analyzed in terms of PQCD.

The matching of certain observables in this transition between a final state involving hadrons and a final state involving jets can be studied by a soft-collinear effective theory (SCET) extracted from PQCD [19] or by specific assumptions such as local parton-hadron duality (LPHD) [20].  The success of PQCD calculations compared to data involving such matching conditions tests the underlying assumption that color confinement and the consequent rearrangement of color in the fragmentation process resulting in a specific configuration of final-state hadrons is dominated by soft, low-momentum transfer processes.  A variation of this phenomenological approach involves QCD-based Monte-Carlo models (such as HERWIG [21], and PYTHIA [22]).  In comparing such models to data, the perturbative scale-evolution of a quark or gluon jet is terminated at some low scale, $\mu_c^2 \cong (1-2) GeV^2$, and the final configuration of hadrons is specified by nonperturbative algorithms motivated by specific assumptions about confinement, such as the Lund String Model [23] or the Cluster Fragmentation Model [24].  The success of this type of data reduction involves the further assumption that it is possible to parameterize the quantum mechanics of the color rearrangement process into simple combinations of probability densities.

One of the most powerful tools in the study of nonperturbative dynamics in QCD involves the numerical simulation non-Abelian fields regularized on a Euclidean lattice.  Lattice QCD has had many significant successes [25,26] but this discrete form of regularization procedure is not well suited to fragmentation dynamics in Minkowski space.  The resulting absence of *ab initio* theoretical information from lattice studies about the important degrees of freedom in the fragmentation processes requires other, more phenomenological approaches to the subject of the type being considered here. The overall topic of fragmentation dynamics requires a very broad focus so it is helpful to concentrate specifically on the mechanisms associated with the spin orientation of a hyperon produced in the final state.

In this sense, it is easy to understand why jets containing $\Lambda_0$ or $\bar{\Lambda}_0$ particles are worthy of special attention. Current understanding of baryon structure in QCD leads to a description of the $\Lambda_0$ based on an isoscalar, $J=0$, $[u,d]$ diquark that is in a $\bar{3}$ representation of SU(3) color and bound to an s quark.[27] The spin orientation of any $\Lambda_0$ is therefore strongly correlated to the spin orientation of the s quark it contains. The fundamental concept of KPR factorization described above then indicates that the dynamical mechanisms leading to the transverse polarization of a $\Lambda_0$ particle must involve soft processes not described by perturbative QCD. The quantum mechanical description of the production mechanism necessarily involves a spin-directed momentum transfer that is <u>odd</u> under a symmetry (here designated $A_\tau$ but sometimes labeled "naïve time reversal" [8]) that can be used to generate idempotent projection operators. These projections ensure that the $A_\tau$-odd dynamics leading to KPR factorization can be described in terms of probability densities. In the discussion below we will model the class of $A_\tau$-odd mechanisms in terms of the nonperturbative production of a virtual $^3P_0$ ($J^{PC}=0^{++}$) $s\bar{s}$ pair with the polarized s quark of the pair being "captured" by the [u,d] diquark. This mechanism is a familiar example of a spin-orbit correlation in QCD that was first discussed systematically by Andersson, Gustafson, Ingelman and Sofstrand [23]. Because both perturbative and non-perturbative components of QCD display an approximate SU(3) flavor symmetry connecting the slightly-more-massive s quark to the light, u and d quarks, related dynamical mechanisms involving the $A_\tau$-odd production of $^3P_0$ $q\bar{q}$ pairs should also occur in the production within QCD jets of the other baryons in the same flavor octet as the $\Lambda_0$, the isodoublets of $(p,n)$ and $(\Xi^0, \Xi^-)$ and the isospin triplet of $(\Sigma^+, \Sigma^0, \Sigma^-)$.

Dramatically large transverse polarization asymmetries $Pd\sigma(pp \Rightarrow \Lambda_0 \uparrow X) = d\sigma(pp \Rightarrow \Lambda_0 \uparrow X) - d\sigma(pp \Rightarrow \Lambda_0 \downarrow X)$ have been observed experimentally [28,29] for $\Lambda_0$'s and other hyperons produced in hadronic collisions. Historically, such asymmetries have been studied in terms of specifically-designed phenomenological models [30,31]. To reconcile these data with QCD, it is important to note that detailed calculations by Dharmaratna and Goldstein [32] have explicitly verified that the mass of the strange quark is not large enough to explain these asymmetries in terms of PQCD processes of the type analyzed by KPR [5] and characterized by expressions similar to Eq. (1). Instead, the large polarization asymmetries for $\Lambda_0$ production require $A_\tau$-odd nonperturbative dynamics. A significant amount of the transverse polarization data for hyperon production involves hyperons found in the "beam fragmentation" or "target fragmentation" region of a baryon and, for these data, it is convenient to use KPR factorization to describe the asymmetries generated by the $A_\tau$- odd mechanisms in terms of <u>polarizing fracture(d) functions</u> and <u>fractured Boer-Mulders functions</u> [33,34] involving the fragmentation of the remnant diquark contained in the beam or target baryon. In contrast, for $\Lambda_0$ hyperons produced in the "current fragmentation" region of a QCD jet, the correlation between the $\Lambda_0$ momentum, spin orientation and jet axis can be defined in terms of the polarizing fragmentation functions included in the Mulders-Tangerman [13] classification of KPR-factorized dynamical mechanisms discussed above. Daniel Boer [35] has used these polarizing fragmentation functions to discuss the process $Pd\sigma(pp \Rightarrow \Lambda_0\uparrow + jetX)$ with the $\Lambda_0$ produced opposite a jet with large transverse momentum. All of these polarization production asymmetries, whether occurring in beam, target or jet fragmentation involve $A_\tau$-odd observables involving the spin-directed momentum $p_{TN}^\Lambda = \vec{p}_\Lambda \cdot (\hat{s}_\Lambda \times \hat{p}_{jet})$ of the $\Lambda_0$ hyperon itself. In contrast, the new type of spin asymmetries proposed in this paper involve a completely separate application of KPR factorization yielding asymmetries of the spin-directed momenta, $p_{TN}^{h_i} = \vec{p}_{h_i} \cdot (\hat{s}_\Lambda \times \hat{p}_{jet})$, for other hadrons, $h_i$, found within the same jet as the transversely polarized $\Lambda_0$. In many ways, these new asymmetries provide more information about the jet fragmentation process than the familiar inclusive $\Lambda_0$ polarization asymmetries.

**Asymmetries for Quantum Number Densities Within QCD Jets**

We can give a simple demonstration of how these new types of asymmetries can occur. Fig. 1 shows a diagram of the momenta for particles in a QCD jet. In this figure, the x-z plane is chosen to be the plane determined by the 3-momentum of the jet, $\vec{P}_{jet}$, and the 3-momentum of the detected $\Lambda_0$, $\vec{p}_\Lambda$. For convenience, the z-axis in this plane is chosen to be along the jet momentum, $\vec{P}_{jet} = |\vec{P}_{jet}|\hat{e}_z$. To complete the orientation of the x-z plane, we look at the weak decay of the $\Lambda_0$, $\Lambda_0 \Rightarrow p\pi^-$, in the $\Lambda_0$ rest frame. For a $\Lambda_0$ with transverse spin in the y-direction, the angular distribution of the decay proton in the $\Lambda_0$ rest frame is given by

$$\frac{dn}{d\Omega} = \frac{1}{4\pi}(1+\alpha \cos\theta_p) \qquad (3)$$

with $\cos\theta_p = (\hat{e}_y \cdot \hat{p}_p)_\Lambda$ and the analyzing power of the weak decay is experimentally determined to be $\alpha \cong 0.642$.[36,37] Therefore, by choosing the orientation of the y-axis for each event containing an observed $\Lambda_0$ by the requirement $\cos\theta_p \geq 0$ we can then specify the x-axis by defining $\hat{x} = \hat{y} \times \hat{z}$. Using the projection operators for $A_\tau$-odd dynamics, we find that the spin density matrix for the production of $\Lambda_0$ particles in the transversity basis for which the Pauli spin matrix $\sigma_y$ is diagonal, we find a spin polarization density for $\Lambda_0$ particles along the positive y-axis of

$$P_\Lambda^y(\theta_p) = \frac{n_y^+ - n_y^-}{n} = \alpha \cos\theta_p \qquad (4)$$

With the constraint $\cos\theta_p \geq 0$, this leads to an ensemble of events with $\Lambda_0$'s polarized in the positive y direction. We will look for spin-directed dynamical asymmetries by measuring the spin-directed momenta $p_{TN}^{h_i} = \vec{p}_{h_i} \cdot (\hat{\sigma}_\Lambda \times \hat{p}_{jet}) = \vec{p}_{h_i} \cdot \hat{e}_x$ for hadrons produced in the neighborhood of the polarized ensemble of hyperons.

The familiar single-spin asymmetry in the production mechanism for $\Lambda_0$ hyperons within the jet of the form $\vec{p}_\Lambda \cdot (\hat{s}_\Lambda \times \hat{P}_{jet})$ is specified in this framework by observing a non-vanishing expectation value for the spin-directed momentum [7], $\langle p_{TN}^\Lambda \rangle$, with $p_{TN}^\Lambda = \vec{p}_\Lambda \cdot \hat{e}_x$ based on the orientation defined above. In the jet fragmentation region, such a non-zero expectation value is conveniently parameterized by the $\Lambda_0$ polarizing fragmentation function, $\Delta^N D_{\Lambda_0\uparrow/q}(z_\Lambda, p_{TN}^\Lambda)$ [11] while in the target fragmentation region it can be parameterized by the polarizing fractured function $\Delta^N M_{\Lambda_0\uparrow/[qq]}^q(x, k_T^2; z_\Lambda, p_{TN}^\Lambda)$ [33,34]. However, for the kinematics of the $\Lambda_0$ particles considered here in the central region of a QCD jet, it is expected that the hyperon whose spin is measured is predominately produced as part of a baryon-antibaryon pair in the jet, and that

$$\langle p_{TN}^\Lambda \rangle = 0. \qquad (5)$$

This assumption should be tested experimentally as part of the measurement of Lambda hadron correlations that is outlined below. If it is not true, there are some nontrivial kinematic correlations between the $\Lambda_0$ momentum and the $p_{TN}$ of the neighboring hadrons that must be considered separately.

To define particle density asymmetries within the QCD jet containing the $\Lambda_0$, it is convenient to parameterize the 4-momentum of the $\Lambda_0$ hyperon

$$P_\Lambda^\mu = (m_T^\Lambda \cosh\eta_\Lambda, p_{TN}^\Lambda \hat{e}_x, m_T^\Lambda \sinh\eta_\Lambda) \qquad (6)$$

in terms of the rapidity variable, $\eta_\Lambda$, and the transverse effective mass $m_T^\Lambda = (m_{\Lambda_0}^2 + p_{TN}^{\Lambda\,2})^{\frac{1}{2}}$. The 4-momenta of neighboring hadrons can then be given

$$P_{h_i}^\mu = (m_T^{h_i} \cosh\eta_i, p_{TN}^{h_i} \hat{e}_x + p_{TS}^{h_i} \hat{e}_y, m_T^{h_i} \sinh\eta_i) \qquad (7)$$

with rapidity $\eta_i$ and transverse effective mass $m_T^{h_i} = (m_{h_i}^2 + p_T^{h_i\,2})^{\frac{1}{2}}$. This pair of 4-momenta defines a set of two-body systems with Mandelstam invariants such as

$$s_{\Lambda h_i} = m_{\Lambda_0}^2 + m_{h_i}^2 + 2m_T^\Lambda m_T^{h_i} \cosh(\eta_i - \eta_\Lambda) - 2 p_{TN}^\Lambda p_{TN}^{h_i}$$

$$t_{\Lambda h_i} = m_{\Lambda_0}^2 + m_{h_i}^2 - 2m_T^\Lambda m_T^{h_i} \cosh(\eta_i - \eta_\Lambda) + 2 p_{TN}^\Lambda p_{TN}^{h_i}$$

for each $h_i$. Because we are interested in particles with small values of these these Mandelstam invariants, particle density distributions can be conveniently defined as functions of $\delta\eta_i = \eta_i - \eta_\Lambda$ and $\delta p_{TN}^i = p_{TN}^{h_i} - p_{TN}^\Lambda$

$$n_i(\delta\eta_i, \delta p_{TN}^i) = \int dp_{TS}^i \frac{d\sigma_{h_i}}{d\eta_i dp_{TN}^i dp_{TS}^i}(\delta\eta_i, \delta p_{TN}^i, p_{TS}^i) \tag{8}$$

For particles in the rapidity interval $\delta\eta_i \in (-\eta_M, +\eta_M)$ in the same jet as the $\Lambda_0$ hyperon, such observables are clearly sensitive to nonperturbative dynamics of the highly virtual system of color SU(3) fields that produce the hadrons. Asymmetries such as

$$\Delta^N n_i(\delta\eta_i, \delta p_{TN}^i) = P_\Lambda^y(\cos\theta_p)[n_i(\delta\eta_i, \delta p_{TN}^i) - n_i(\delta\eta_i, -\delta p_{TN}^i)] \tag{9}$$

for specific hadronic species such as $h_i = K^+, \pi^+, \bar{p},...$ are characteristic of spin-orbit dynamics in the formation of the corresponding two-hadron systems during the hadronization stage of the fragmentation process. In this expression, the form of the Polarization density for the $\Lambda_0$ hyperon is given by Eq. (4). In this form, a large number of events containing hyperons with "measured" polarization can be combined to form density distributions containing many hadrons. Summing over the hadrons in the event with the density distributions for identified hadrons weighted by conserved quantum numbers such as electric charges, $Q_i$, strangeness, $S_i$ or baryon number $B_i$, then give local quantum number asymmetries

$$\Delta^N Q(\delta\eta, \delta p_{TN}) : \Delta^N S(\delta\eta, \delta p_{TN}) : \Delta^N B(\delta\eta, \delta p_{TN})$$

that provide markers labeling the quantum numbers involved in these coherent, nonperturbative effects. We note that it is also interesting to use polarization weighted rapidity asymmetries such as

$$\Delta^\eta n_i(\delta\eta_i, \delta p_{TN}^i) = P_\Lambda^y(\cos\theta_p)[n_i(\delta\eta_i, \delta p_{TN}^i) - n_i(-\delta\eta_i, \delta p_{TN}^i)] \tag{10}$$

in addition to unweighted rapidity asymmetries, such as $\delta n_i(\delta\eta) = [n_i(\delta\eta) - n_i(-\delta\eta)]$, in combination with the spin-directed asymmetries of (9) in order to study the quantum number flow for the coherent dynamics of the virtual systems involved in the evolution of the fragmentation process. At this point it is instructive to consider some simple examples of spin-directed quantum number density asymmetries.

**Examples of Fragmentation Mechanisms**

As mentioned in the introduction, the splendid successes of PQCD [2,3] have provided crucial tools for exploring the large-momentum, small-distance, frontier of the Standard Model. However, in order to explain the motivation for studying observables involving KPR factorized asymmetries in the transverse momenta and rapidities of particles in a QCD jet containing a transversely polarized $\Lambda_0$, it is instructive to consider an expanded space-time picture of the fragmentation process for a segment of a QCD jet that can lead to such asymmetries. This is because the confinement of SU(3) color necessarily plays an integral role in those dynamical mechanisms leading to $A_\tau$-odd transverse spin observables and KPR factorization. The momentum-space formulation of the Feynman rules for PQCD required for the study of "hard-scattering" factorization" [1] can provide the twist-expansion for classifying $A_\tau$-odd observables [12] but a study of the origin of virtual SU(3)-colored subsystems with $|\langle L \rangle| \geq 1$ within a larger volume of confined SU(3)- colored fields is most conveniently framed in a space-time formulation of quantum field theory. The sketch in Fig. 2, thus, indicates a region of space behind a quark or gluon traveling with large momentum the positive z –direction. Because of the confining properties of the color force, it is assumed that all color fields are restricted to a cylindrical region and it is further assumed that an internal segment of that cylindrical region is uniformly expanding. These assumptions can be given a quantum field-theoretical interpretation by casting them in terms of expectation values for observables formed from the color fields. For example, the assumption of uniform expansion can be expressed by specifying that, in a Lorentz frame co-moving with the center of the internal segment (as indicated by CS in the diagram), hadrons formed only from fields confined to the right of CS will have positive values for the z-component of their momenta, $\langle p_z^{h_R} \rangle \geq 0$ and that hadrons formed only from the color fields to the left of CS will have negative values for the z-component of their momenta, $\langle p_z^{h_L} \rangle \leq 0$.

The time ordering of "events" in an extended system can, of course, depend on the Lorentz frame from which the "events" are viewed. The isolation of an SU(3) color singlet state with the quantum numbers of a $\Lambda_0$ hyperon and with measurable 4-momentum and spin orientation requires a sequence of such events. Two separate dynamical mechanisms for producing a $\Lambda_0$ particle are shown from the CS frame as indicated by the sketches in Figs. 3 and 4. The formation of a $\Lambda_0 \uparrow$ with spin directed in $+\hat{e}_y$ direction from a $J=0$ [u,d] diquark and a polarized s quark from a virtual $^3P_0$ $s-\bar{s}$ pair with $\langle L_y = -1 \rangle$ is crudely indicated by sequences shown in Fig. 3. In the CS-co-moving frame, the location of the [u,d] diquark in the first scene is to the left of center. The non-local expression,

$$\langle L_y \rangle = \langle zp_x - xp_z \rangle = -1 \qquad (11)$$

for the virtual s-$\bar{s}$ pair is indicated in the sketch. The consequent "weedeater" annihilation of a portion of the color flux is suggested by the configuration in the second scene of this sketch. The process results in hadronic cluster including the polarized $\Lambda_0$, $[X_L\Lambda_0 \uparrow]$ with $\langle p_z \rangle \leq 0; \langle p_x \rangle \geq 0$ separated from the color fields of a second hadronic cluster containing the $\bar{s}$ quark, $[\bar{s}X_R]$ that has $\langle p_z \rangle \geq 0; \langle p_x \rangle \leq 0$ in the CS reference frame. These sketches in Fig. 3 therefore provide a partial representation of the familiar "string-breaking" mechanism for color confinement found in the Lund fragmentation model [24]. This semi-classical picture shown here has also been used by Artru, Czyzewski, and Yabuki [38] to model the rank-1 "favored" and rank-2 "unfavored" Collins functions for pseudoscalar meson production. When viewed from the CS frame, the relative kinematics of the distinct hadronic clusters produced by this sequence are independent of other dynamical considerations.

The sketches in Fig. 4 follow the sequence of Fig. 3 with one crucial difference. In the drawings of Fig. 4, the virtual [u,d] diquark is located to the right of center in the co-moving CS frame. Note that this change in the location of the annihilated flux results in a separation of clusters with the $[\Lambda_0 \uparrow X_R]$ cluster having $\langle p_z \rangle \geq 0; \langle p_x \rangle \leq 0$ while the $[X_L \bar{s}]$ cluster now has $\langle p_z \rangle \leq 0; \langle p_x \rangle \geq 0$ in the CS Lorentz frame. Note that the combination of the processes in Fig. 3 and Fig. 4 will result in the prediction $\langle p_{TN}^\Lambda \rangle = 0$ for a $\Lambda_0$ polarized in the positive y direction (as given in Eq. (5)) unless one of the orderings is favored over the other. In the central region of a QCD jet, there is no reason for such a preference to occur. Of course, for Lorentz boosts along the z-axis, a boost-invariant description of the result dynamical sequence can be given by expressing the z-components of hadron momenta in terms of rapidity and the x-component of hadron momenta in terms of the spin-directed transverse moment, $p_{TN}^h$. Both $\delta\eta_i$ and $\delta p_{TN}^i$ are preserved under such boosts and the rapidity-space density asymmetries described by eqs. (9) and (10) provide significant information about the mechanisms involved in determining the spin orientation of the $\Lambda_0$ hyperon. From Fig. 3 we can draw the inference that hadrons in the jet sharing the quantum numbers of the $\bar{s}$ quark $\left(S = +1, Q = +\frac{1}{3}, B = -\frac{1}{3}\right)$ with $\delta\eta = \eta_i - \eta_\Lambda$ positive will preferentially have $\delta p_{TN} = \delta p_{TN}^i - \delta p_{TN}^\Lambda$ negative. Correspondingly, Fig. 4 shows that hadrons with these quantum numbers and $\delta\eta$ negative will tend to have $\delta p_{TN}$ positive. The figures demonstrate the possible existence of interesting quantum number density asymmetries oriented by the spin of the $\Lambda_0$ hyperon.

Two general, simplifying, assumptions about the properties of the fragmentation process that are illustrated in the crude drawings of Figs. 2-4 should be mentioned.

1. The figures incorporate the assumption that color confinement in QCD plays an important role both in restricting the overall framework and in the final resolution of the virtual state into specific, isolated, color-singlet hadrons. It is by no means certain that they incorporate all of the constraints imposed on confined systems of SU(3) color.
2. The sequences shown incorporate the suggestion inferred from causal arguments in quantum theory that jet fragmentation is semi-local and exothermic in the sense that the non-Abelian color flows in the processes that form localized color-singlet clusters release some of the energy and momentum stored in coherent field configurations to provide momentum kicks to the emerging hadrons.

One particular mechanism is definitely absent in this these sketches. The figures do not show the complicated topological structure required at the boundaries of confined regions of color flux in gauge theories. The existence of such quantum structures, however, can be shown to restrict the mechanisms that can contribute to exothermic flux-breaking dynamical processes.[39]

The simple examples illustrated here leave a clear signal in the strangeness density asymmetry, $\Delta^N S(\delta\eta, \delta p_{TN})$. A sketch showing a naïve calculation for this asymmetry is shown in Fig. 5. The relationship between this asymmetry and the closely related asymmetries, $\Delta^N Q(\delta\eta, \delta p_{TN})$ and $\Delta^N B(\delta\eta, \delta p_{TN})$ for charge and baryon number require additional dynamical assumptions concerning the flavor combinations for mesons and antibaryons even within the framework of these simple examples. Because of the different combinations of mesons and antibaryons that can be formed with the $\bar{s}$ quark, the tendency is for these asymmetries to have longer correlation lengths in rapidity space. It is important to keep in mind that other virtual systems with nonzero orbital angular momentum can also contribute to the spin-orientation of an s-quark in the hadronization stage of the fragmentation process and they would leave behind different patterns of quantum number density asymmetries. The examples shown here imply that the experimental comparison of the different quantum number density distributions can provide significant, direct, insight into the $A_\tau$-odd dynamics of vitual hadronic systems including the detected $\Lambda_0$.

**Experimental Considerations**

The quantum number density asymmetries discussed above can be measured in any hard-scattering process that involves QCD jets of sufficient energy to include baryon-antibaryon pairs containing $\Lambda_0$ or $\bar{\Lambda}_0$ hyperons. The experimental constraints on different types of measurements that are required for the study of such asymmetries need to be carefully examined in the context of specific detectors. Basically, the complete set of requirements involve measurements that:

1. Accurately determine the jet axis, $\hat{z} = \hat{P}_{jet}$, for a QCD jet containing a $\Lambda_0$ hyperon.
2. Measure the 3-momentum, $\vec{p}_\Lambda$, of the specific $\Lambda_0$ reconstructed from the weak decay $\Lambda_0 \Rightarrow p\pi^-$. The transverse component of this momentum relative to the jet axis must be nonzero in order to define an x-z plane. It is necessary to orient the x-z plane containing $\vec{p}_\Lambda$ and $\hat{P}_{jet}$ by specifying that the component of the momentum for the decay proton normal to this x-z plane is directed along the <u>positive</u> y-axis and use this information to give $\hat{e}_x = +\hat{e}_y \times \hat{e}_z$.
3. Measure the decay angle, $\theta_p$, of the proton in the rest frame of the $\Lambda_0$ to determine the transverse polarization density, $P_\Lambda^y(\cos\theta_p) \cong 0.642\cos\theta_p$ with the constraint $\cos\theta_p \geq 0$.
4. Take advantage of particle identification and momentum resolution for other hadrons within the jet to measure rapidities and transverse momenta for particles with rapidities near $\eta_\Lambda$.

Each of these requirements involves explicit considerations of acceptances, accuracy and systematic errors. For example, the effect of the precession of the $\Lambda_0$ spin around the magnetic field within the detectors before the weak decay depends on the magnitude and orientation of $\vec{p}_\Lambda$ with respect to these fields. This precession needs to be considered carefully before comparing jets of different orientations within the detector. The author is not prepared to evaluate all such issues at this time and, thus, is forced to rely more on hope than on detailed knowledge of the experimental capabilities of the appropriate detectors in order to advocate for such measurements. The hope is that measurement of the quantum number asymmetries discussed here can, indeed, performed and will consequently provide tools to enhance the study of QCD jets first begun by Field and Feynman [40] . It is notable that a recent paper by Quigg [41] suggests that examination of 3-dimensional plots of individual events in $\eta \otimes \vec{p}_T$ space from high-energy colliders could uncover patterns leading to unexpected insights. The techniques suggested here merely use $\Lambda_0$ spin measurements to orient local sections for a subset of such plots so that they can be combined to provide information about coherent subprocesses in QCD. It might also be possible, if other patterns emerge, that experimenters will find ways to use the oriented plots in a more creative way.

One particular set of possible systematic errors concerning spin states needs to be discussed. Start with an ensemble of $\Lambda_0$'s with two different spin states $n_+(0)$ and $n_-(0)$ with the spin quantization axis chosen to be in the $\hat{y}$ direction. We define

$$P_\Lambda^y(0) = \frac{n_+(0) - n_-(0)}{n_+(0) + n_-(0)} \qquad (12)$$

and normalize to $n_+(0) + n_-(0) = 1$. Based on eq. (3), if we measure $\cos\theta_p \geq 0$ for the proton in the weak decays of the $\Lambda_0$'s we will find

$$n_+(\theta_p) = n_+(0)[1 + \alpha \cos\theta_p]$$
$$n_-(\theta_p) = n_-(0)[1 - \alpha \cos\theta_p] \tag{13}$$

This leads to

$$n_+(\theta_p) + n_-(\theta_p) = 1 + P_\Lambda^y(0)\cos\theta_p$$
$$n_+(\theta_p) - n_-(\theta_p) = P_\Lambda^y(0) + \alpha\cos\theta_p \tag{14}$$

If we also detect decays with protons at an angle $\pi - \theta_p$ where $\cos(\pi - \theta_p) = -\cos\theta_p$ we can use the identities

$$n_+(\pi - \theta_p) = n_-(\theta_p)$$
$$n_-(\pi - \theta_p) = n_+(\theta_p) \tag{15}$$

to show that

$$P_\Lambda^y(\cos\theta_p) = \frac{[n_+(\theta_p) - n_-(\theta_p)] - [n_+(\pi - \theta_p) - n_-(\pi - \theta_p)]}{[n_+(\theta_p) + n_-(\theta_p)] + [n_+(\pi - \theta_p) + n_-(\pi - \theta_p)]} = \alpha\cos\theta_p \tag{16}$$

These simple manipulations verify that the arguments in the text regarding the definition of the y- axis used to orient the x-z plane for the spin-dependent density asymmetries lead an unbiased definition of the polarization density as given in Eq (4).
.

**Discussion**

  The quantum number density asymmetries discussed here present a new type of transverse spin observable that can be measured in any process where QCD jets can be found containing hyperons or anti-hyperons with spin-analyzing weak decays. The interpretation of the resulting spin-oriented asymmetry measurements is simplest for $\Lambda_0$'s or $\bar\Lambda_0$'s produced in the central rapidity region of the jet but measurements in other kinematic regimes can also produce interesting information. Simple arguments can be used to show that the spin-orbit dynamics generating the asymmetries extends over a space-time region that plays an important role in the hadronization phase of the fragmentation process for those hadrons with rapidities near that of the hyperon with measured transverse spin. The simple model involving the virtual creation of an $s\bar{s}$ pair in a $^3P_0$ configuration with $J^{PC} = 0^{++}$ that was used to illustrate to role of orbital angular momentum in an expanding system with confined color fields provides an interesting starting point for considering these new spin-dependent observables.

  The experimental challenges for measuring such quantum number density asymmetries are significant. However, the hope expressed here is that this type of transverse spin observables probing the interior dynamics of QCD jets can be found in high-energy collisions providing information about spin-dependent dynamics in nonperturbative QCD without the requirement for Siberian Snakes to produce polarized beams and without the insertion of polarized targets.

## Acknowledgements

The author is extremely grateful for the series of detailed and penetrating questions posed to him by Homer Neal about the dynamical mechanisms leading to $\Lambda_0$ polarization in high-energy scattering processes. Those questions led directly to the initial draft of this manuscript. An insightful reading of the initial manuscript by the referee from Phys. Rev. D generated strong suggestions for changes necessary to provide additional clarity. Detailed instructive comments from Homer Neal, Daniel Scheirich and Gary Goldstein have also provided invaluable guidance for shaping the discussion above.

**Figure Captions**

Fig. 1. This drawing shows the momenta of particles making up a QCD jet with its total momentum directed along the z-axis, $\vec{P}_{jet} = |\vec{P}_{jet}|\hat{e}_z$. The jet is assumed to contain a $\Lambda_0$ particle with momentum $\vec{p}_\Lambda$ in the x-z plane. The spin of the $\Lambda_0$ is quantized transverse to the x-z plane and the positive direction for the y-axis for this quantization is chosen by specifying that the weak decay $\Lambda_0 \Rightarrow p\pi^-$ results in a proton momentum with $\vec{p}_p \cdot \hat{e}_y \geq 0$. (In a frame with $\vec{p}_p \cdot \hat{e}_y \leq 0$, all the 3-momenta in the event, including $\vec{p}_\Lambda$ are to be rotated around the z axis by $\pi$ to meet this requirement. This procedure orients the x-z plane so that $\hat{e}_x = +(\hat{e}_y \times \hat{e}_z)$ enables the study of coherent, spin-directed dynamical mechanisms. The result of Kane, Pumplin and Repko, [5] guarantees that asymmetries in $p_{TN}^{h_i}$ for other hadrons in the event cannot be the result of PQCD processes.

The inset of this figure defines the angle, $\theta_p$, that in the rest frame of the $\Lambda_0$, determines the decay asymmetry for the weak decay. A measurement of $\theta_p$ gives the spin polarization density $P_y(\theta_p) = \alpha \cos\theta_p$ with $\alpha = 0.642 \pm 0.013$. [36,37]

Fig. 2 This sketch indicates a cylindrically symmetric system behind a QCD constitutent (quark or gluon) produced with original momentum $\vec{P}_{const} \cong |\vec{P}_{jet}|\hat{e}_z$ in a hard-scattering event. As the system of SU(3) color fields trailing this constituent expands along the z axis, we assume there exists a segment of the cylinder between the discs A and E where the local properties of the system can be approximately described by two order parameters, $\Phi(\rho,z,t) = \langle G_{\mu\nu}^a G^{\mu\nu a} \rangle$ (a density with $0^{++}$ quantum numbers) and $\Pi(\rho,z,t) = \langle {}^*G_{\mu\nu}^a G^{\mu\nu a} \rangle$ ( a density with $0^{-+}$ quantum numbers) and that these densities vanish for $\rho \geq \rho_{MAX}$. It is further assumed that this segment of the overall system is uniformly expanding and that, in a Lorentz frame co-moving with the color-averaged density in the center of the segment (indicated by CS in the sketch), momentum

conservation gives $\langle p_z \rangle_{AB}^{CS} \leq 0$ for the cylindrical subsystem $\Box AB \Box$ bounded by the discs labeled A,B and $\langle p_z \rangle_{DE}^{CS} \geq 0$ for the cylindrical subsystem $\Box DE \Box$ bounded by the discs D,E. These assumptions will be used to constrain momentum observables for color-singlet hadrons produced in the fragmentation process.

Fig. 3   Two drawings indicating the sequence for an example mechanism involving the creation of an $s\bar{s}$ pair in a $^3P_0$ state to the right of a $[ud]$ diquark to produce a polarized $\Lambda_0$ within a QCD jet. The sequence suggests the formation of a hadronic cluster with the quantum numbers of an $\bar{s}$ quark with $\delta\eta \geq 0$ and $\delta p_{TN} \leq 0$.

Fig 4   Two drawings indicating the same sequence as Fig. 3 except that the $s\bar{s}$ pair is created to the left of the $[ud]$ diquark. This configuration leads to a cluster containing the $\bar{s}$ quark with $\delta\eta \leq 0$ and $\delta p_{TN} \geq 0$.

Fig. 5   The sample mechanisms in Figs. 3-4 are used to calculate naïve estimates for the strangeness density asymmetry $\Delta^N S(\delta\eta, \delta p_{TN})$, the charge density asymmetry $\Delta^N Q(\delta\eta, \delta p_{TN})$, and the baryon number density asymmetry $\Delta^N B(\delta\eta, \delta p_{TN})$ for $\delta p_{TN} = 0.1$ GeV/c. The calculations assume that the hadron closest to the $\Lambda_0$ in the fragmentation process contains the $\bar{s}$ quark. The connection between the correlation lengths in rapidity space for the asymmetries $\delta Q(\delta\eta, \delta p_{TN})$ and $\delta B(\delta\eta, \delta p_{TN})$ involve more dynamical assumptions.

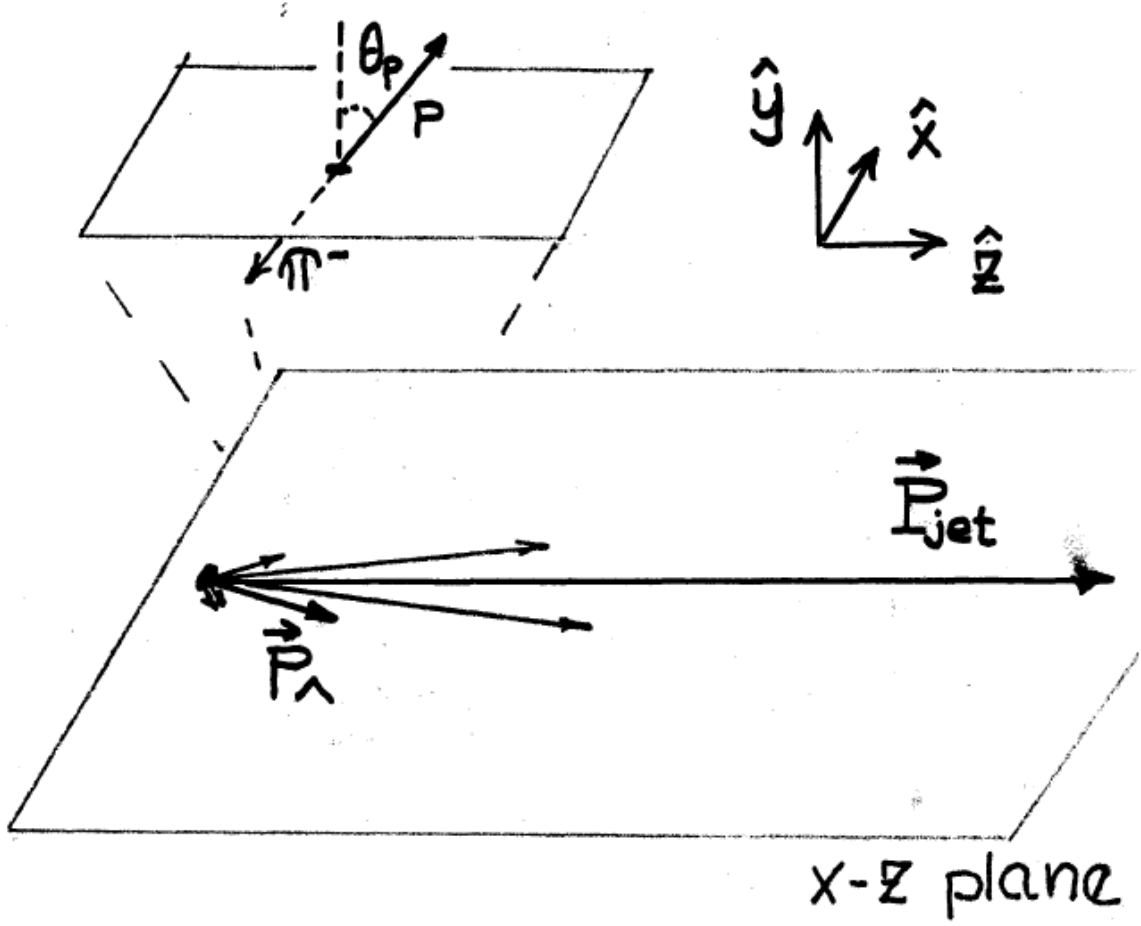

Figure 1

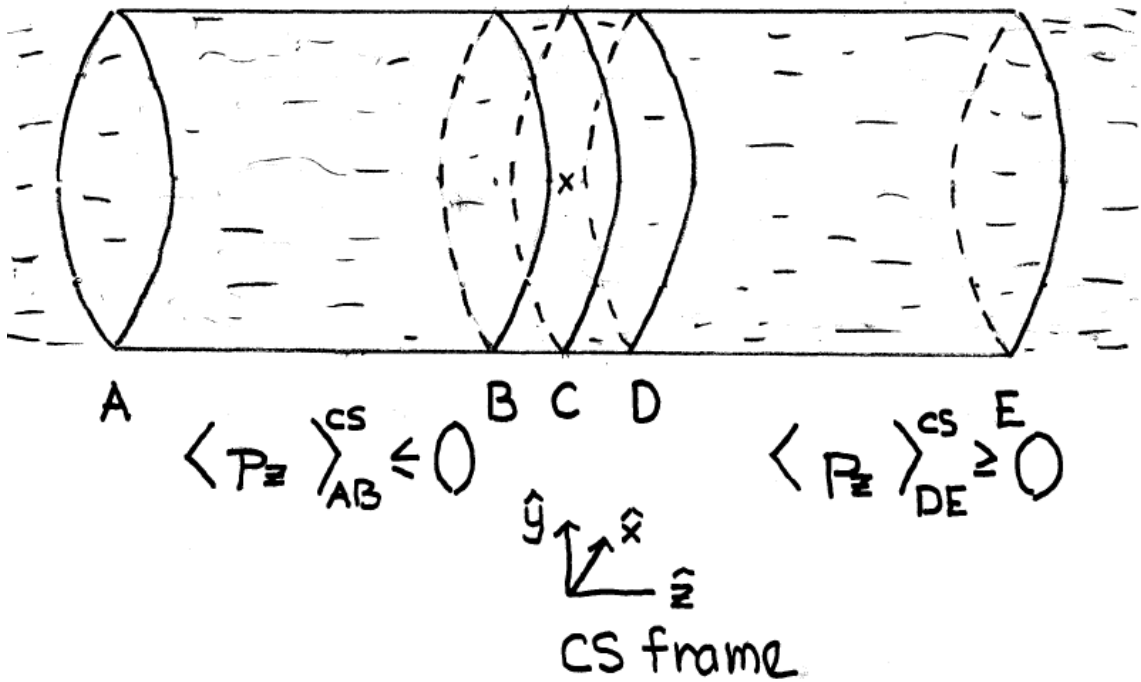

Figure 2

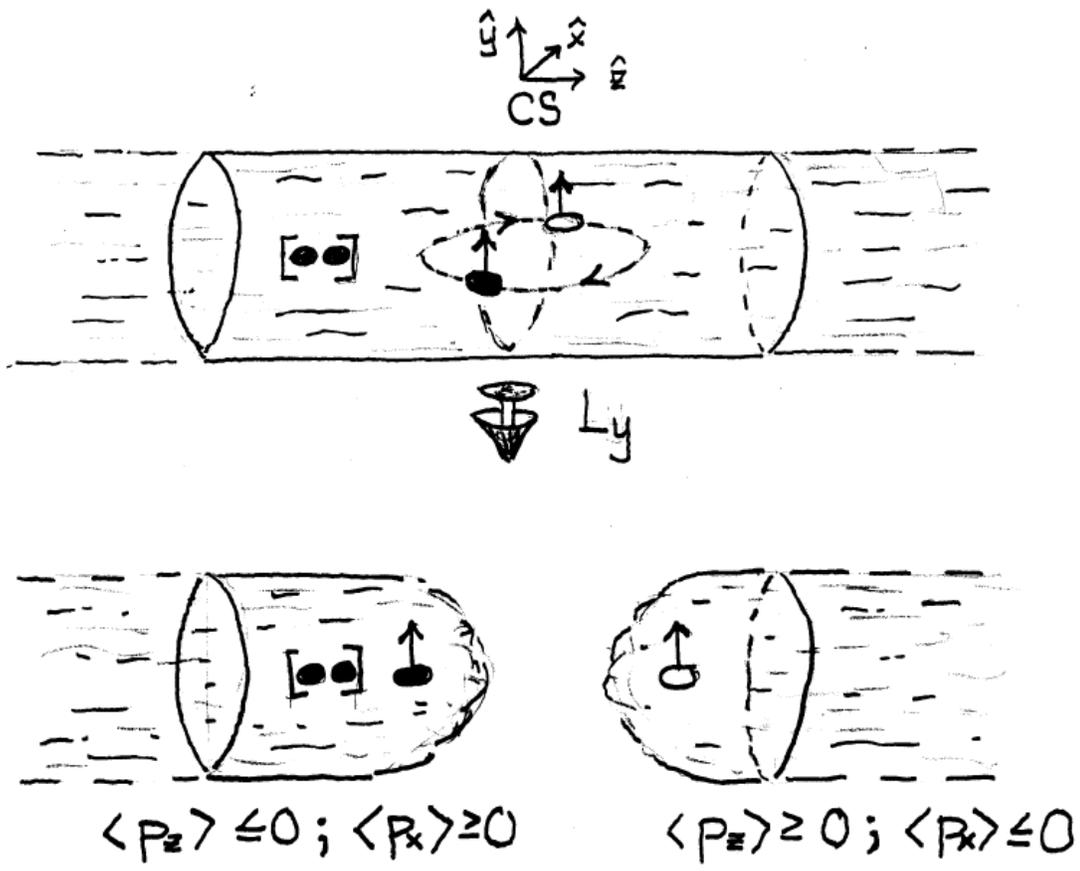

Figure 3

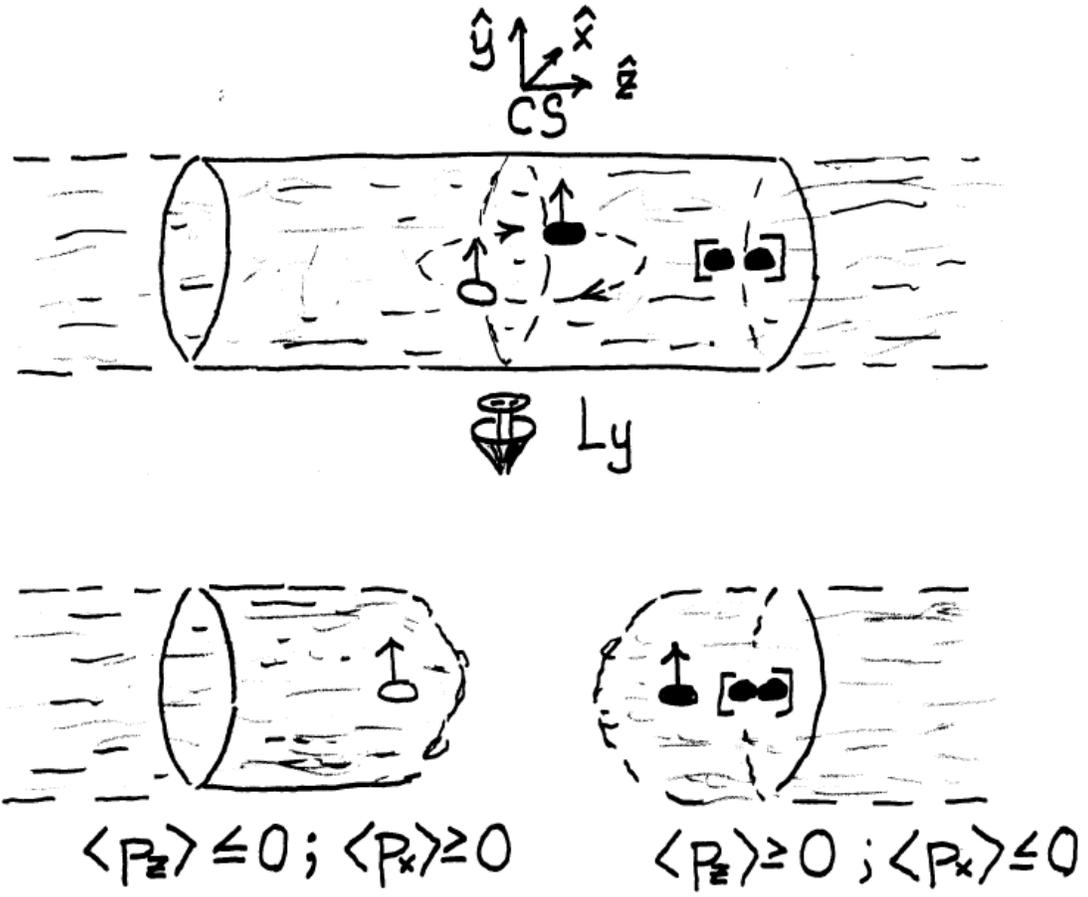

Figure 4

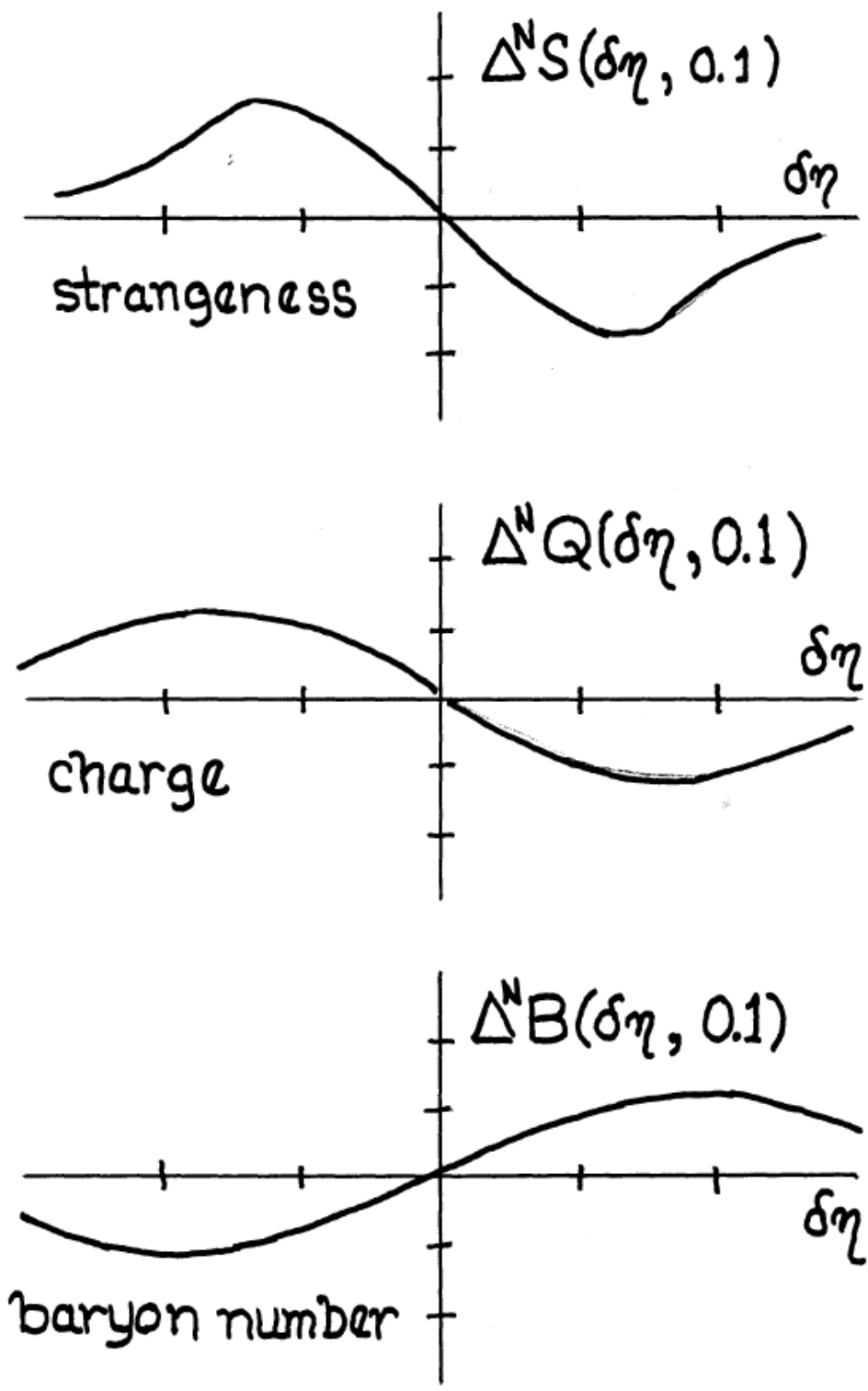

Figure 5